\documentclass[12pt,a4paper]{article}
\usepackage{amsmath,epsfig}
\begin{document}
\begin{titlepage} 
\vspace*{0.5cm}
\begin{center}
{\Large\bf Virial expansion and TBA in $O(N)$ sigma models}
\end{center}
\vspace{2.5cm}
\begin{center}
{\large J\'anos Balog and \'Arp\'ad Heged\H us}
\end{center}
\bigskip
\begin{center}
Research Institute for Particle and Nuclear Physics,\\
Hungarian Academy of Sciences,\\
H-1525 Budapest 114, P.O.B. 49, Hungary\\ 
\end{center}
\vspace{3.2cm}
\begin{abstract}
We study the free energy of the $1+1$ dimensional $O(N)$ nonlinear 
$\sigma$-models for even $N$ using the TBA equations proposed recently. 
We give explicit formulae for the constant solution of the TBA equations
(Y-system) and calculate the first two virial coefficients. The free
energy is also compared to the leading large $N$ result.
\end{abstract}

\end{titlepage}

\newcommand{\be}{\begin{equation}}
\newcommand{\bea}{\begin{eqnarray}}

\newcommand{\Eps}{\Epsilon}
\newcommand{\gM}{\mathcal{M}}
\newcommand{\dD}{\mathcal{D}}
\newcommand{\gG}{\mathcal{G}}
\newcommand{\pa}{\partial}
\newcommand{\eps}{\epsilon}
\newcommand{\La}{\Lambda}
\newcommand{\De}{\Delta}
\newcommand{\nonu}{\nonumber}
\newcommand{\beq}{\begin{eqnarray}}
\newcommand{\eeq}{\end{eqnarray}}
\newcommand{\ka}{\kappa}
\newcommand{\an}{\ensuremath{\alpha_0}}
\newcommand{\bn}{\ensuremath{\beta_0}}
\newcommand{\dn}{\ensuremath{\delta_0}}
\newcommand{\al}{\alpha}
\newcommand{\bm}{\begin{multline}}
\newcommand{\fm}{\end{multline}}
\newcommand{\de}{\delta}

\section{Introduction}

A large class of integrable two-dimensional models can be
described as perturbed conformal field theories (PCFT). If the
perturbing relevant operator is chosen so that 
perturbation preserves integrability then the spectrum of
the resulting (in general massive) integrable model and
its exact S-matrix can be conjectured \cite{Z1}. 
Further, the set of TBA equations \cite{Z2} describing the 
ground state energy in finite volume (or equivalently the 
free energy of the system at finite temperature $T$) can be 
written down.

Non-linear $\sigma$-models (NLS) are another important class 
of integrable models. Their spectrum and S-matrix have
been obtained by bootstrap methods \cite{ZZ} but the 
corresponding TBA system was not known until recently. In 
\cite{Fendley,Fendley2} (based on earlier constructions
\cite{FatZam,Zamo,ABC}) NLS models are represented as
limits of PCFT models. More concretely, the NLS model
whose fields take value in the coset space $G/H$ can be
described as the $k\rightarrow\infty$ limit of the
$(G/H)_k$ conformal coset model \cite{GKO} perturbed by
a certain relevant operator. This way of representing
NLS models gives additional insight into the structure
of the model and is reminescent of the description of the
model as the infinite flavour limit of a certain fermionic
model \cite{PolWieg}. Moreover, it allows one to calculate
the ground state energy of the NLS model by considering the 
limiting case of the TBA system, available for
finite values of the Kac-Moody level~$k$.

Because of the obvious importance of the NLS models in this
paper we perform a number of tests for the above TBA system.
For simplicity, we restrict our attention here to the $O(N)$
NLS models for even $N$. We show that in the 
$T\rightarrow\infty$ limit the correct UV central charge
is reproduced. In the opposite (low temperature or
equivalently low density) limit there exists a systematic
expansion, the virial expansion. It is known \cite{Ma}
that the virial coefficients are completely determined by
the scattering data alone. This is useful since the virial
coefficients can also be calculated from the TBA equations 
by solving the linearized system in Fourier space. We work out
the first two virial coefficients directly from the
S-matrix and also from the TBA system and find that they
agree completely. Finally, NLS models can also be solved 
in the $1/N$ expansion. We calculate the free energy in the 
leading large $N$ limit and compare it to the 
$N\rightarrow\infty$ limit of the virial coefficients 
calculated previously.

\section{TBA equations}

Here we study the TBA system proposed in \cite{Fendley} for the $O(N)$
nonlinear sigma model for the case of even $N$.
The TBA equations are formulated in terms of the densities 
$\varepsilon_0(\theta)$ and $\varepsilon^{(a)}_m(\theta)$, where the
index $a$ labels the nodes of the $D_r$ Dynkin diagram ($N/2=r\geq2$)
and the range of the lower index is $m=1,2,\dots,k-1$ ($k\geq2$). 
Let us introduce
\bea
A^{(a)}_m(\theta)&=&\frac{g}{4\pi}\int_{-\infty}^\infty
\frac{d\theta^\prime}{\cosh
\frac{g}{2}(\theta-\theta^\prime)}\ln\left[1+
e^{-\varepsilon^{(a)}_m(\theta^\prime)}\right],\label{Aam}\\
A_0(\theta)&=&\frac{g}{4\pi}\int_{-\infty}^\infty
\frac{d\theta^\prime}{\cosh
\frac{g}{2}(\theta-\theta^\prime)}\ln\left[1+
e^{-\varepsilon_0(\theta^\prime)}\right],\\
B^{(a)}_m(\theta)&=&\frac{g}{4\pi}\int_{-\infty}^\infty
\frac{d\theta^\prime}{\cosh
\frac{g}{2}(\theta-\theta^\prime)}\ln\left[1+
e^{\varepsilon^{(a)}_m(\theta^\prime)}\right],\label{Bam}
\end{eqnarray}
where (\ref{Aam}) and (\ref{Bam}) are defined for all $a=1,2,\dots r$
and $m=1,2,\dots,k-1$. By convention we take 
$B^{(a)}_0(\theta)=B^{(a)}_k(\theta)=0$. $g=2r-2$ is the Coxeter number of
$D_r$. We further define
\be
\psi_{ab}(\theta)=\int_{-\infty}^\infty\,\mathrm{d}\omega\,e^{i\omega\theta}
{\cal N}_{ab}(\omega),
\label{psiab}
\end{equation}
where for $r\geq3$ ${\cal N}_{ab}(\omega)$ is the matrix inverse of
\be
{\cal M}_{ab}(\omega)=2\cosh\left(\frac{\pi\omega}{g}\right)\delta_{ab}
-I_{ab}.
\end{equation}
Here $I_{ab}$ is the incidence matrix of the $D_r$ Dynkin diagram.

For later use we note that ${\cal N}_{a1}(\omega)$ is given by
\be
{\cal N}_{a1}(\omega)=\frac{q^a+q^{2r-2-a}}{1+q^{2r-2}} \quad a=1,2,\dots,
r-2
\label{Na1I}
\end{equation}
and
\be
{\cal N}_{r-1,1}(\omega)={\cal N}_{r1}(\omega)
=\frac{q^{r-1}}{1+q^{2r-2}},
\label{Na1II}
\end{equation}
where $q=e^{-\frac{\pi\vert\omega\vert}{g}}$.
We will assume that (\ref{Na1II}) holds also 
for $r=2$, by definition in this case.

The TBA equations proposed in \cite{Fendley} are
\be
\varepsilon_0(\theta)=\frac{1}{T}(M\cosh\theta-H)-\frac{1}{2\pi}\sum_{a=1}^{r} 
\int_{-\infty}^{\infty} \mathrm{d}\theta^\prime
\psi_{a1}(\theta-\theta^\prime)\ln\left[1+
e^{\varepsilon_{1}^{(a)}(\theta^\prime)} \right]
\label{TBA1}
\end{equation}
and for $a=1,2,\dots,r$ and $m=1,2,\dots k-1$
\be
\varepsilon_{m}^{(a)}(\theta)=A_0(\theta) 
( \delta_{a1}\delta_{m1}+\delta_{a2}\delta_{m1}\delta_{r2} )
+B^{(a)}_{m+1}(\theta)+B^{(a)}_{m-1}(\theta)
-\sum_{b=1}^{r} I_{ab}A_{m}^{(b)}(\theta).
\label{TBA2}
\end{equation}
In (\ref{TBA1}) $M$ is the mass of the particles and $H$ is a chemical
potential coupled to particle number.

Finally the pressure of this one-dimensional gas is given by
\be
p=\frac{MT}{2\pi}\int_{-\infty}^\infty\mathrm{d}\theta\,\cosh\theta
\ln\left[1+e^{-\varepsilon_0(\theta)}\right].
\label{pressure}
\end{equation}

The above TBA system describes the perturbed $(O(N)/O(N-1))_k$ coset
model for finite $k$. The pressure of the $O(N)$ NLS model can be obtained 
from it by taking the $k\rightarrow\infty$ limit \cite{Fendley}.

\section{Virial expansion}

We have introduced the chemical potential $H$ to facilitate the virial
expansion of the pressure. In the limit $H\rightarrow-\infty$ the densities
can be expanded in powers of the fugacity $y=e^{H/T}$ as
\bea
\varepsilon_0(\theta)&=&\frac{1}{T}(M\cosh\theta-H)+\sum_{\mu=0}^\infty
\varepsilon_0^{(\mu)}(\theta)y^\mu,\label{virialI}\\
\varepsilon_m^{(a)}(\theta)&=&\sum_{\mu=0}^\infty
\varepsilon_m^{(a)(\mu)}(\theta)y^\mu,\label{virialII}
\end{eqnarray}
whereas the pressure is expanded as
\be
p=T\sum_{\mu=1}^\infty p^{(\mu)} y^\mu.
\label{virialexp}
\end{equation}
Alternatively, one can get the same expansion coefficients (without
introducing the chemical potential) by considering the low temperature
expansion of the pressure. In this case the coefficient $p^{(\mu)}$
is of order $\exp(-\mu M/T)$.

It is known \cite{Ma} that for the case of a system of $N$ particles 
(independently of integrability) the virial expansion coefficients
are determined by the S-matrix. If all particles have the same mass $M$
then the first two virial coefficients are
\be
p^{(1)}=\frac{n}{2\pi}\int_{-\infty}^\infty\mathrm{d}\theta\,M
\cosh\theta\,e^{-\frac{M}{T}\cosh\theta}
\label{virial1}
\end{equation}
and $p^{(2)}=p^{(2)}_1+p^{(2)}_2$, where
\be
p^{(2)}_1=-\frac{ns}{4\pi}\int_{-\infty}^\infty\mathrm{d}
\theta\,M
\cosh\theta\,e^{-\frac{2M}{T}\cosh\theta}
\label{virial21}
\end{equation}
and
\be
p^{(2)}_2=\frac{1}{4\pi^2}\sum_l
\int_{-\infty}^\infty\mathrm{d}\theta
\int_{-\infty}^\infty\mathrm{d}\theta^\prime\,M\cosh\theta\,
\phi_l(\theta-\theta^\prime)
e^{-\frac{M}{T}(\cosh\theta+\cosh\theta^\prime)}.
\label{virial22}
\end{equation}
Here $n=N$, 
$\phi_l(\theta)=-i\frac{\mathrm{d}}{\mathrm{d}\theta}\ln S_l(\theta)$,
where $\{S_l(\theta);\ l=1,2,\dots,N^2\}$
 are the eigenvalues of the
2-particle S-matrix and $s$ is a factor related to particle statistics.
For the case of diagonal scattering, $s=\mp1$ for bosonic/fermionic
type particles \cite{KM}, whereas for all known non-diagonal cases
(where all particles are of fermionic type) $s=N$.
Finally, for later use, we introduce the Fourier transform
\be
\sum_l\phi_l(\theta)=\int_{-\infty}^\infty\mathrm{d}\omega\,
e^{-i\omega\theta}\,\tilde\phi(\omega).
\label{tildephi}
\end{equation}
For the $O(N)$ nonlinear sigma model, using the eigenvalues of the
bootstrap S-matrix \cite{ZZ} we get
\be
\tilde\phi(\omega)=N^2\,\frac{e^{-\pi\vert\omega\vert}+
e^{-\frac{2\pi\vert
\omega\vert}{N-2}}}{1+e^{-\pi\vert\omega\vert}}-e^{-\pi\vert\omega\vert}-
\frac{N^2-N+2}{2}\,\,e^{-\frac{2\pi\vert\omega\vert}{N-2}}.
\label{ONtildephi}
\end{equation}

\section{Solution of the Y-system}

Let us consider the generic $r\geq3$ case first.
The leading terms in the expansion (\ref{virialII}) turn out to be
$\theta$-independent and the constants
\be
\hat x^{(a)}_m=e^{\varepsilon^{(a)(0)}_m(\theta)}
\label{xdef}
\end{equation}
can be determined by solving the equations
\be
\left(\hat x^{(a)}_m\right)^2\prod_{b=1}^r\left[1+\frac{1}{\hat x^{(b)}_m}
\right]^{I_{ab}}=\left(1+\hat x^{(a)}_{m+1}\right)
\left(1+\hat x^{(a)}_{m-1}\right)
\label{hatxsystem}
\end{equation}
for $a=1,2,\dots,r$, $m=1,2,\dots,k-1$,
where by convention $\hat x^{(a)}_0=\hat x^{(a)}_k=0$.
This ubiquitous equation is called the constant Y-system and its
solution is well known for all classical Dynkin diagrams \cite{Kuniba}.
It is usually solved by relating it to the Q-system equations
\be
\left(Q^{(a)}_m\right)^2=Q^{(a)}_{m+1}Q^{(a)}_{m-1}
+\prod_{b=1}^r\left[Q^{(b)}_m\right]^{I_{ab}}
\label{Qsystem}
\end{equation}
for $a=1,2,\dots,r$, $m=1,2,\dots,k-1$,
with \lq\lq boundary condition''
\be
Q^{(a)}_{0}=Q^{(a)}_{k}=1.
\label{Qboundary}
\end{equation}
We are interested in positive solutions of (\ref{Qsystem}),
related to (\ref{hatxsystem}) by
\be
\hat x^{(a)}_m=\frac{Q^{(a)}_{m+1}Q^{(a)}_{m-1}}
{\prod\limits_{b=1}^r\left[Q^{(b)}_m\right]^{I_{ab}}}.
\label{hatxQ}
\end{equation}

We will need the explicit solution of (\ref{Qsystem})
 only for $a=1$.
This can be written as \cite{Kuniba}
\be
Q^{(1)}_m=\frac{p(m+r-1)}{p(r-1)}\prod_{j=1}^{2r-3}\frac{p(m+j)}{p(j)},
\label{Q1}
\end{equation}
with $p(n)=\sin\frac{n\pi}{k+2r-2}$.
In particular,
\be
Q^{(1)}_1=\frac{p(r)p(2r-2)}{p(1)p(r-1)}\qquad
{\rm and}\qquad
Q^{(1)}_2=\frac{p(r+1)p(2r-1)p(2r-2)}{p(1)p(2)p(r-1)}.
\label{Q112}
\end{equation}

We can now calculate the first virial coefficient. It is of the form
(\ref{virial1}), where the coefficient $n$ can be expressed
in terms of the constants (\ref{xdef}) as
\be
n=\prod_{a=1}^r\left[1+\hat x^{(a)}_1\right]^{{\cal N}_{a1}(0)}.
\label{nI}
\end{equation}

If we use (\ref{Na1I}), (\ref{Na1II}) together
with (\ref{hatxQ}), (\ref{nI}) can be written as
\be
n=\prod_{a=1}^{r-1}\left[1+\hat x^{(a)}_1\right]=Q^{(1)}_1.
\label{nII}
\end{equation}
From (\ref{Q112}) we see that 
$\lim\limits_{k\rightarrow\infty}\, n=2r=N$. 

\section{The second virial coefficient}

Encouraged by the agreement of the first virial coefficient
with the expected form we now go one step further and
calculate the second virial coefficient. 
Using the expansions
(\ref{virialI}) and (\ref{virialII}) in (\ref{TBAI}),
(\ref{TBAII}) and going to Fourier space we find that
$p^{(2)}_1$ is of the form (\ref{virial21}) with $s=n$ and
$p^{(2)}_2$ is of the form (\ref{virial22}) with
\be
\tilde\phi(\omega)=n^2\sum_{a=1}^r{\cal N}_{a1}(\omega)
S^{(a)}_1(q).
\label{Sa1}
\end{equation}
Here the functions $S^{(a)}_1(q)$ can be calculated by solving the
linear equations 
\be
\left(q+q^{-1}\right)\left(1+\frac{1}{\hat x^{(a)}_m}\right)
S^{(a)}_m=\delta_{a1}\delta_{m1}+S^{(a)}_{m+1}+S^{(a)}_{m-1}+
\sum_{b=1}^rI_{ab}\frac{1}{\hat x^{(b)}_m}S^{(b)}_m
\label{Sam}
\end{equation}
for $a=1,2,\dots,r$, \ $m=1,2,\dots, k-1$, where
$S^{(a)}_0=S^{(a)}_k=0$.

The solution of the linear equations (\ref{Sam}) can be given in 
terms of the numbers $Q^{(a)}_m$ satisfying (\ref{Qsystem}). 
Although their explicit expression
is known \cite{Kuniba}, it is actually easier to verify 
the validity of the solution given below by using the 
Q-system equations (\ref{Qsystem}) directly.

The solution for $S^{(a)}_m(q)$ is given as follows.
Using the definitions
\be
b^{(a)}_m=\frac{Q^{(a)}_m}{Q^{(a)}_{m-1}}\qquad
m=1,2,\dots,k
\label{bam}
\end{equation}
and
\bea
I^{(1)}_m&=&b^{(1)}_{m+1},\\
I^{(a)}_m&=&\frac{b^{(a)}_{m+1}}{b^{(a-1)}_m},
\quad a=2,\dots,r-2\\
I^{(r-1)}_m&=&
\frac{b^{(r-1)}_{m+1}b^{(r-1)}_m}{b^{(r-2)}_m}+
\frac{b^{(r-1)}_{m+1}}{b^{(r-1)}_{m-1}},\label{omit}\\
I^{(r)}_m&=&0
\end{eqnarray}
for $m=1,2,\dots,k-1$ (for $m=1$ the second term in (\ref{omit}) is
absent) together with $I^{(a)}_0=I^{(a)}_k=0$
we first define
\bea
H^{(a)}_m(q)&=&I^{(a)}_m\,q^{m+a-1}-I^{(a+1)}_m\,q^{m+a+1},
\quad a=1,2,\dots,r-1\\
H^{(r)}_m(q)&=&H^{(r-1)}_m(q)
\end{eqnarray}
for $m=0,1,\dots,k$.
Next we define
\be
\hat S^{(a)}_m(q)=\frac{1}{n}\left[H^{(a)}_m(q)-q^{k+g}
H^{(a)}_{k-m}\left(\frac{1}{q}\right)\right].
\end{equation}
Note that both $H^{(a)}_m(q)$ and $\hat S^{(a)}_m(q)$
are finite polynomials in $q$.

The solution of (\ref{Sam}) is finally given by
\be
S^{(a)}_m(q)=\frac{1}{1-q^{2k+2g}}\left[
\hat S^{(a)}_m(q)-q^{2k+2g}\hat S^{(a)}_m\left(\frac{1}{q}
\right)\right].
\label{solution}
\end{equation}

Using the solution (\ref{solution}) we can calculate
\be
\tilde\phi(\omega)=\frac{1}{1+q^g}\frac{1}{1-q^{2k+2g}}
\left[\hat K(q)-q^{2k+3g}\hat K\left(\frac{1}{q}\right)
\right],
\label{final}
\end{equation}
where
\be
\hat K(q)=n^2\sum_{a=1}^{r-1}\left(q^a+q^{g-a}\right)
\hat S^{(a)}_1(q).
\end{equation}
Putting everything together we have
\be
\hat K(q)=
Q^{(1)}_2\,q^2+(n^2-1)q^g-\left(n^2-Q^{(1)}_2\right)
q^{g+2}-q^{2g}
\end{equation}
and using also 
\be
\lim\limits_{k\rightarrow\infty} Q^{(1)}_2=
(r+1)(2r-1)=\frac{N^2+N-2}{2}
\label{Q12}
\end{equation}
we see that (\ref{final})
indeed reproduces (\ref{ONtildephi}) in the limit
$k\rightarrow\infty$.

\section{Calculation of the UV central charge}

For this purpose it turns out
to be useful to rewrite the TBA equations a little.
Let us introduce the new variables
\bea
\epsilon_0(\theta)&=&\varepsilon_0(\theta),\qquad\ \ \ \ \ \  
L_0(\theta)=\ln\left[1+e^{-\epsilon_0(\theta)}\right],\\
\epsilon^{(a)}_m(\theta)&=&-\varepsilon^{(a)}_m(\theta),\qquad
L^{(a)}_m(\theta)=\ln\left[1+e^{-\epsilon^{(a)}_m(\theta)}
\right].
\end{eqnarray}
In terms of these new variables the TBA equations take the
standard form
\bea
\epsilon_0&=&\frac{1}{T}\left(M\cosh\theta-H\right)-
\sum_{a=1}^r\psi_{a1}*L^{(a)}_1,\label{TBAI}\\
\epsilon^{(a)}_m&=&\sum_{b,c=1}^r\psi_{ab}I_{bc}*L^{(c)}_m-
\psi_{a1}\delta_{m1}*L_0-\sum_{b=1}^r\psi_{ab}*\left(
L^{(b)}_{m+1}+L^{(b)}_{m-1}\right).\label{TBAII}
\end{eqnarray}
Our convention for convolution is
\be
\left(f*g\right)(\theta)=\frac{1}{2\pi}\int_{-\infty}^\infty
\mathrm{d}\theta^\prime f(\theta-\theta^\prime)g(\theta^\prime)
\end{equation}
and $L^{(a)}_0(\theta)=L^{(a)}_k(\theta)=0$ by convention.

We note that the physical particles correspond to $\epsilon_0$.
They are of mass $M$ and are coupled to the chemical potential
$H$. All other excitations corresponding to the densities 
$\epsilon^{(a)}_m$ are \lq\lq magnons''. They are massless 
and are not coupled to the chemical potential.

The standard form of the TBA equations allows us to calculate the
UV central charge the standard way \cite{KM}. It can be written as 
$c_{\rm UV}=c^{(1)}-c^{(2)}$ with
\be
c^{(1)}=\frac{6}{\pi^2}\left\{
{\cal L}\left(\frac{x_0}{1+x_0}\right)+
\sum_{a=1}^r\sum_{m=1}^{k-1}{\cal L}\left(
\frac{x^{(a)}_m}{1+x^{(a)}_m}\right)
\right\}
\label{c1}
\end{equation}
and
\be
c^{(2)}=\frac{6}{\pi^2}
\sum_{a=1}^r\sum_{m=1}^{k-1}{\cal L}\left(\frac{\hat x^{(a)}_m}{1+
\hat x^{(a)}_m}\right),
\label{c2}
\end{equation}
where ${\cal L}(z)$ is Roger's dilogarithm. In (\ref{c2}) the set of numbers
$\{\hat x^{(a)}_m\}$ satisfy (\ref{hatxsystem})
and are the same as occuring in the virial expansion.
They can be found in \cite{Kuniba} where not only their values,
but also the corresponding
sum of dilogarithms is evaluated (conjectured):
\be
c^{(2)}=\frac{k(k-1)r}{k+2r-2}.
\label{c2value}
\end{equation}

The numbers $\{x^{(a)}_m\}$ are defined as the solution of a similar
problem. They satisfy the equations
\be
\left(x^{(a)}_m\right)^2\prod_{b=1}^r\left[1+\frac{1}{x^{(b)}_m}
\right]^{I_{ab}}=\left(1+x^{(a)}_{m+1}\right)\left(1+x^{(a)}_{m-1}\right)
\label{xsystem}
\end{equation}
for $a=1,2,\dots,r$, $m=1,2,\dots,k-1$ with \lq\lq boundary conditions''
\be
x^{(a)}_k=0,\qquad\qquad x^{(a)}_0=\delta_{a1}x_0,
\label{xbc}
\end{equation}
where
\be
x_0=\prod_{a=1}^r\left(1+x^{(a)}_1\right)^{{\cal N}_{a1}(0)}.
\label{x0}
\end{equation}
 
Similarly to the previous case, it is also useful to introduce the
\lq\lq Q-system'' here. We denote the corresponding numbers by 
$\{R^{(a)}_m\}$ and define for $m=0,1,\dots,k$
\be
x^{(a)}_m=\frac{R^{(a)}_{m+1}R^{(a)}_{m-1}}
{\prod\limits_{b=1}^r\left[R^{(b)}_m\right]^{I_{ab}}}.
\label{xR}
\end{equation}
The numbers $R^{(a)}_m$ are assumed to be nonnegative 
and satisfy the Q-system equations
\be
\left(R^{(a)}_m\right)^2=R^{(a)}_{m+1}R^{(a)}_{m-1}
+\prod_{b=1}^r\left[R^{(b)}_m\right]^{I_{ab}}
\label{Rsystem}
\end{equation}
for $a=1,2,\dots,r$, $m=0,1,\dots,k$ with boundary condition
\be
R^{(a)}_{k+1}=0,\qquad\qquad R^{(a)}_{-1}=\delta_{a1}
\prod_{b=1}^r\left(R^{(b)}_0\right)^{I_{1b}}.
\label{Rbc}
\end{equation}

We have found the solution of the modified Q-system (\ref{Rsystem}),
(\ref{Rbc}). The components corresponding to the fork part of the
Dynkin diagram can be written as
\be
R^{(r-1)}_m=R^{(r)}_m=\prod_{\alpha\in\Delta_+}
\frac{\sin\frac{\langle\alpha\vert(k-m)\omega_r+\rho\rangle\,\pi}{k+g-1}}
{\sin\frac{\langle\alpha\vert\rho\rangle\,\pi}{k+g-1}}
\prod_{l=0}^{r-2}
\frac{\sin\frac{(k-m+2l+1)\pi}{2(k+g-1)}}
{\sin\frac{(2l+1)\pi}{2(k+g-1)}}
\frac{\sin\frac{(2l+1)\pi}{k+g-1}}
{\sin\frac{(k-m+2l+1)\pi}{k+g-1}}.
\label{fork}
\end{equation}
Here $\Delta_+$ is the set of positive roots, $\rho$ is half the sum
of the positive roots, $\omega_r$ is the $r^{\rm th}$ fundamental weight
and scalar product is normalized in weight space so that $\langle\alpha
\vert\alpha\rangle=2$ for the roots $\alpha$. For $m=0$ the ratio
\be
\frac{\sin\frac{(k-m+g-1)\pi}{k+g-1}}
{\sin\frac{(g-1)\pi}{k+g-1}}
\end{equation}
has to be omitted from the first factor in (\ref{fork}) together with the 
simultaneous omission of its inverse from the last factor.

The remaining components, $\{R^{(a)}_m {\rm \ for\ } a=r-2,\dots,2,1\}$, can be
calculated from the equations (\ref{Rsystem}) with $a=r-1,\dots,3,2$. 
We have verified that the last Q-system equation (for $a=1$) is then
also satisfied.

Based on extensive numerical study, we conjecture that if we use the 
solution of the \lq\lq modified Y-system'' (\ref{xsystem}), (\ref{xbc}) 
obtained this way in the dilogarithm sum (\ref{c1}) we get
\be
c^{(1)}=\frac{k(k-1)r+k(r-1)}{k+2r-3}.
\label{c1value}
\end{equation}

Putting (\ref{c1value}) and (\ref{c2value}) together, we see that
$c_{\rm UV}$ can also be written in the suggestive form
\be
c_{\rm UV}=c(k,N)-c(k,N-1),
\end{equation}
where 
\be
c(k,N)=\frac{k\frac{N(N-1)}{2}}{k+N-2}
\label{sugN}
\end{equation}
is the central charge of the $O(N)$ WZNW conformal model at level $k$.

\section{The $O(4)$ model}

The $r=2$ case is somewhat special and we discuss it separately. From 
(\ref{psiab}) and (\ref{Na1II}) it follows that in this case
\be
\psi_{a1}(\theta)=\frac{1}{\cosh\theta}=\psi(\theta).
\end{equation}

We now define the new variables
\be
\epsilon_0(\theta)=\varepsilon_0(\theta),\qquad\qquad
\epsilon_m(\theta)=-\varepsilon^{(1)}_m(\theta),\qquad\qquad
\epsilon_{-m}(\theta)=-\varepsilon^{(2)}_m(\theta).
\end{equation}
We further define
\be
l_m(\theta)=\ln\left[1+e^{-\epsilon_m(\theta)}\right]
\end{equation}
for $m=1-k,\dots,0,\dots,k-1$ and $l_{\pm k}(\theta)=0$.
This allows us to write the TBA equations (\ref{TBA1}) and
(\ref{TBA2}) in the compact form
\be
\epsilon_m+\psi*(l_{m+1}+l_{m-1})=\frac{1}{T}(M\cosh\theta-H)
\delta_{m0}
\label{TBAr2}
\end{equation}
for $m=1-k,\dots,0,\dots,k-1$.

In this form it is easy to recognize that (\ref{TBAr2})
is the standard TBA system \cite{Zamo} associated to the $A_{2k-1}$
Dynkin diagram with $m=0$ corresponding to the single 
massive node. This observation allows us to 
determine the UV central charge immediately from
\be
c_{\rm UV}=c(A_{2k-1})-2\,c(A_{k-1}).
\end{equation}
Using the well-known formula \cite{Kuniba}
\be
c(A_l)=\frac{l(l+1)}{l+3}
\end{equation}
we get
\be
c_{\rm UV}=\frac{3k^2}{(k+1)(k+2)},
\end{equation}
which is the expected result since it can also be written as
\be
c_{\rm UV}=c(k,4)-c(2k,3).
\label{c43}
\end{equation}
Note that the case of the $O(4)/O(3)$ sigma model is
exceptional in that the KM level corresponding to O(3) is
doubled. For $N=3$ (\ref{sugN}) is not valid either, we
have $c(k,3)=3k/(k+2)$ instead.

The Y-system equations in this case simplify to
\be
\hat x_m^2=(1+\hat x_{m+1})(1+\hat x_{m-1})
\label{hatxr2}
\end{equation}
for $m=1,2,\dots,k-1$ and $\hat x_0=\hat x_k=0$. 
They are solved in terms of the Q-sytem as
\be
\hat x_m=Q_{m+1}Q_{m-1},\qquad{\rm where}\qquad
Q_m^2=1+Q_{m+1}Q_{m-1}.
\end{equation}
The Q-system has a simple solution in this case. It is
\be
Q_m=\frac{p(m+1)}{p(1)}.
\end{equation}
It is easy to calculate the virial coefficients $p^{(1)}$
and $p^{(2)}_1$. They are of the form (\ref{virial1})
and (\ref{virial21}) respectively with
\be
s=n=(1+\hat x_1)=Q_1^2,
\end{equation}
which indeed reproduces $N=4$ in the $k\rightarrow\infty$
limit.

The Fourier transform (\ref{tildephi}) entering the second
virial coefficient in this case is
\be
\tilde\phi(\omega)=\frac{2n^2q}{1+q^2}\,S_1(q),
\label{tildephir2}
\end{equation}
where $S_1(q)$ is determined by solving 
\be
\left(q+\frac{1}{q}\right)\left(1+\frac{1}{\hat x_m}\right)
S_m(q)=\delta_{m1}+S_{m+1}(q)+S_{m-1}(q)
\label{Sr2}
\end{equation}
for $m=1,2,\dots,k-1$ with boundary condition $S_0(q)=S_k(q)=0$.
The solution of (\ref{Sr2}) is
\be
S_m(q)=\frac{1}{Q_1}\,\frac{1}{1-q^{2k+4}}\left[
\hat S_m(q)-q^{2k+4}\hat S_m\left(\frac{1}{q}\right)\right]
\end{equation}
for $m=1,2,\dots,k-1$, where
\be
\hat S_m(q)=\frac{Q_{m+1}}{Q_m}\,q^m-\frac{Q_{m-1}}{Q_m}\,q^{m+2}.
\end{equation}
Finally we find
\be
\tilde\phi(\omega)=\frac{2nq^2}{1+q^2}\,\frac{1}{1-q^{2k+4}}\left[
Q_2-q^2+q^{2k}(1-Q_2q^2)\right],
\end{equation}
which reduces to the expected result
\be
\tilde\phi(\omega)=\frac{8q^2(3-q^2)}{1+q^2}
\end{equation}
in the $k\rightarrow\infty$ limit.

\section{Large $N$ limit}

Finally, we consider the $N\rightarrow\infty$ limit,
where the $O(N)$ $\sigma$-models are exactly solvable.
This limit was studied in \cite{Luscher}, where
the exact formula
\be
\ln\frac{M(R)}{M}=F(z)=2\sum_{m=1}^\infty K_0(mz)
+{\cal O}\left(\frac{1}{N}\right)
\label{MR}
\end{equation}
for the mass gap $M(R)$ in finite volume (corresponding to
a circle of radius $R$) was found. Here $K_0$ is a modified 
Bessel function and $z=M(R)R$, which can be obtained,
as a function of $\zeta=MR$, by solving this implicit equation.

Using similar techniques, we calculated the $N\rightarrow
\infty$ limit of the pressure. This requires the calculation
of the ground state energy in finite volume \cite{KM}.
We find
\be
p(R)=\frac{N}{8\pi}\left\{M^2-M^2(R)\right\}
+\frac{N}{2\pi}M^2(R)\sum_{m=1}^\infty K_2(mz)
+{\cal O}(1),
\label{Np}
\end{equation}
where $K_2$ is a modified Bessel function and the temperature
is given by $T=1/R$.

We first study the $R\rightarrow0$ limit of the effective
central charge
\be
\tilde c(R)=\frac{6R^2}{\pi}p(R).
\end{equation}
For short distances $F(z)\sim \pi/z$ implying \cite{Luscher}
\be
z\sim\frac{\pi}{\ln\left(\frac{1}{\zeta}\right)}
\end{equation}
consistently with AF perturbation theory. This can be used
to determine the short distance (or high temperature) limit
of the effective central charge. We find
\be
\tilde c(R)=N\left\{1-
\frac{3}{2\ln\left(\frac{1}{\zeta}\right)}
+{\cal O}\left(\frac{1}{\ln^2\left(\frac{1}{\zeta}\right)}
\right)\right\}+{\cal O}(1),
\end{equation}
as expected.

In the opposite ($R\rightarrow\infty$) limit (\ref{MR})
can be expanded as
\be
\frac{M(R)}{M}=1+\sum_{\nu=1}^\infty \Delta_\nu(\zeta),
\label{Delta}
\end{equation}
where $\Delta_\nu(\zeta)={\cal O}(e^{-\nu\zeta})$. The first 
two terms of this expansion are
\be
\Delta_1(\zeta)=2K_0(\zeta)\qquad {\rm and}
\qquad \Delta_2(\zeta)=2K_0^2(\zeta)+
2K_0(2\zeta)-4\zeta K_0(\zeta)K_1(\zeta).
\label{Delta12}
\end{equation}
Similarly, the pressure (\ref{Np}) can be expanded as
\be
p(R)=\frac{1}{R}\sum_{\nu=1}^\infty p^{(\nu)}(\zeta),
\label{pexp}
\end{equation}
where $p^{(\nu)}(\zeta)={\cal O}(e^{-\nu\zeta})$. As explained
above, these expansion coefficients must be the same as the 
virial coefficients of (\ref{virialexp}). Using (\ref{Delta12})
in the expansion of (\ref{Np}), for the first two expansion
coefficients we get  
\bea
p^{(1)}(\zeta)&=&\frac{NM}{\pi}K_1(\zeta)+{\cal O}(1),
\label{p1}\\
p^{(2)}(\zeta)&=&\frac{NM}{2\pi}\left[
K_1(2\zeta)-2\zeta K_0^2(\zeta)\right]+{\cal O}(1).
\label{p2}
\end{eqnarray}

The agreement of (\ref{p1}) and (\ref{p2}) with the first
two virial coefficients of (\ref{virialexp}) in the
large $N$ limit is obvious for $p^{(1)}$, since (\ref{p1})
and (\ref{virial1}) are exactly the same for any model
(independently of the interaction). It is not so obvious for
the second coefficients. First one has to note that the
(constant) ${\cal O}(N^2)$ term of (\ref{ONtildephi})
exactly cancels (\ref{virial21}). There remains the 
${\cal O}(N)$ term
\be
N\left(\frac{1}{2}+\pi\vert\omega\vert-
\frac{2\pi\vert\omega\vert}{1+e^{-\pi\vert\omega\vert}}
\right),
\end{equation}
which, after Fourier transformation, indeed reproduces 
(\ref{p2}).

\vspace{1cm}
{\tt Acknowledgements}

\noindent 
We thank P. Fendley for a correspondence and for suggesting us to study
the large $N$ limit. This investigation was supported in part by the 
Hungarian National Science Fund OTKA (under T030099, T029802 and T034299).

\end{document}